\documentclass[runningheads]{llncs}
\usepackage[T1]{fontenc}
\usepackage{float}
\usepackage{graphicx}
\usepackage{amsmath}
\usepackage{subcaption}
\usepackage{hyperref}
\usepackage{url}
\usepackage{color}

\begin{document}
\title{Hybrid Monitoring for Early Fault Detection in Cloud-Native 5G Systems}
\titlerunning{Hybrid Monitoring for Early Fault Detection in Cloud-Native 5G Systems}

\author{Anton Andersson\inst{1,2}, Sai Akshara Naineni\inst{1,2}, Mats Jansborg\inst{2}, Yixing Zhang\inst{1}, Romaric Duvignau\inst{1}}

\authorrunning{A. Andersson, S. A. Naineni et al.}

\institute{Department of Computer Science and Engineering, Chalmers University of Technology and University of Gothenburg, SE-412 96 Gothenburg, Sweden\\
\email{\{antonand,akshara\}@student.chalmers.se}, \email{\{yixing,duvignau\}@chalmers.se}
\and
Ericsson AB, Gothenburg, Sweden\\
\email{mats.jansborg@ericsson.com}
}
\maketitle              
\begin{abstract}
This paper presents the design, implementation, and evaluation of NetMon, a hybrid network monitoring system designed for Kubernetes-based 5G packet core deployments, specifically evaluated on Ericsson’s Access and Mobility Management Function (AMF) clusters. NetMon combines eBPF-based passive kernel-level traffic observation with active TCP probing and centralized correlation to detect and localize network degradation within seconds. The evaluation results demonstrate that the system detects faults as subtle as 10ms of added latency or 5\% packet loss, correctly attributes them to the affected infrastructure component, and maintains this capability under application loads up to 50\% simulated UE load. The total resource overhead of 3.4 millicores CPU and 4.5 MiB memory per pod suggests that the approach is promising for further validation without impacting the monitored workload. The hybrid approach addresses a gap in existing monitoring tools: standard health checks cannot detect partial degradation, scrape-based systems introduce detection delays measured in tens of seconds, and purely passive tools cannot verify idle network paths. By combining these complementary techniques and centralizing the analysis, the system provides the early detection and fault localization capabilities required for maintaining service quality in cloud-native 5G infrastructure.

\end{abstract}

\section{Introduction}

Monitoring distributed systems is challenging because interacting components fail in subtle, non-obvious ways. As systems scale, determining whether issues originate in application logic, shared resources, or the network becomes increasingly difficult~\cite{bailis2014network}. These challenges are amplified in cloud-native 5G core networks, where microservice-based network functions deployed on Kubernetes create dynamic communication paths where small degradations propagate across components~\cite{ericsson_cloud_native_5g}. Early indicators such as higher jitter, slower tail latency, and dropped packets appear before visible faults.

The 5G core network is functionally decoupled into two main structural layers: a control plane that handles administrative and state logic, and a user plane that processes data packet forwarding~\cite{nunziati_monitoring_2025}. 

The control plane comprises three primary service functions that interact via service-based interfaces (SBIs):
\begin{itemize}
    \item The \textbf{Access and Mobility Management Function (AMF)} manages terminal registration, authentication, and mobility tracking across radio nodes.
    \item The \textbf{Session Management Function (SMF)} handles network address allocation, session establishment, and user plane node policy enforcement.
    \item The \textbf{Policy Control Function (PCF)} delivers routing, Quality of Service (QoS), and subscriber-specific policy rules to the control operations.
\end{itemize}

In contrast, the user plane is represented by the \textbf{User Plane Function (UPF)}, which acts as the data-plane anchor routing traffic between the Radio Access Network (RAN) and the external Data Networks (DN).

Crucially, the control plane and the user plane interact via well-defined reference points rather than inline service busses. One of these points of interaction is the \textbf{N4 interface}, which forms a direct reference link between the SMF and the UPF. Through this interface, the SMF programmatically pushes Packet Forwarding Control Protocol (PFCP) rules down to the UPF, mapping session-state logic to raw data plane instructions. In our monitoring framework, verifying the integrity and latency of this inter-plane communication path is paramount, as impairments along the N4 interface directly degrade session handling and user-plane forwarding performance across the 5G core infrastructure.


\subsection{Related Work}
Existing approaches leave significant gaps. Active probing does not scale well~\cite{partitioning_thesis_2022}. Metrics-based tools (e.g., Prometheus) operate on large scrape intervals, missing transient degradation. Logging is reactive. Distributed tracing misses kernel-level issues. Current AMF monitoring relies mostly on liveness checks that cannot detect early-stage degradation within inter-microservice communication paths, and passive telemetry alone struggles to distinguish genuine degradation from normal variation~\cite{ibidunmoye2015performance}.

The core problem is therefore detecting subtle network degradation such as latency shifts, jitter increases, and intermittent packet loss, before it cascades into visible failures, while keeping overhead acceptable for production 5G deployments. This paper presents \emph{NetMon}, a hybrid tool that addresses this gap by combining passive eBPF-based kernel-level observation with active TCP probing between pods~\cite{soldani_ebpf_2023}. Lightweight agents on each pod probe peers and collect kernel-measured RTT, jitter, retransmission rates, and TCP anomalies, reporting to a central correlator that identifies failure patterns and localizes their source without modifying application code~\cite{rice2023learning}.

\subsection{Research Questions}
\label{sec:research questions}

The research is guided by the following question:

\begin{itemize}
    \item How can early-stage network degradation in cloud-native 5G systems be detected and localized using a lightweight hybrid approach combining passive kernel-level monitoring with active probing?
\end{itemize}

Supported by sub-question:
\begin{itemize}
    \item How can independently observed anomalies be correlated to distinguish pod-level faults, worker issues, and partitions?
\end{itemize}

\subsection{Contributions}
\label{sec:contributions}

To answer these research questions, we present the design, implementation, and evaluation of NetMon, delivering the following concrete contributions:

\begin{itemize}
    \item \textbf{Hybrid monitoring architecture}: To address our primary research question, we design a hybrid monitoring architecture that combines eBPF passive telemetry with active TCP probing, enabling early fault detection through a general-purpose Kubernetes tool evaluated on a 5G AMF deployment.

    \item \textbf{Cluster-wide correlation engine}: To address the correlation of distributed anomalies, we develop an engine that aggregates local reports to identify failure patterns such as peer down, latency source, worker isolation, inter-node failure, and pod isolation thereby distinguishing pod-level issues from infrastructure-level faults.

    \item \textbf{Visualization pipeline}: To ensure operator visibility, we implement a monitoring pipeline integrated with Prometheus and a custom dashboard to export per-pod and per-link health metrics in real time.

    \item \textbf{Live evaluation}: To validate our approach under realistic conditions, we conduct a live evaluation of detection timeliness, accuracy, and resource overhead on a multi-worker AMF cluster, successfully extending monitoring coverage to previously unmonitored internal communication paths.
\end{itemize}

\section{System Design}

This section presents the design of NetMon, including its overall architecture, distributed monitoring agents, local anomaly-detection mechanisms, and centralized correlation process. To ensure the system is viable for production 5G environments, the architecture is engineered to fulfill four primary design objectives:

\begin{itemize}
    \item \textbf{Early detection}: NetMon aims to detect network degradation within seconds, identifying infrastructure faults before standard application-level health checks surface visible problems~\cite{huang2017gray}.
    \item \textbf{Lightweight operation}: The monitoring components are designed to run with a minimal resource footprint to avoid interfering with production workloads under high stress.
    \item \textbf{Cluster-wide correlation}: The architecture isolates faults by distinguishing localized pod-level failures from worker-node issues and broader network partitions.
    \item \textbf{Transparent deployment}: The system is deployed seamlessly as sidecar containers without requiring any modifications to the monitored application's source code.
\end{itemize}

\subsection{Architecture Overview}

The system follows a distributed architecture with two components: agents deployed as sidecars on each monitored pod, and a central server running on the controller pod.
\vspace{-0.5cm}

\begin{figure}[htbp]
    \centering
    \includegraphics[width=0.5\linewidth]{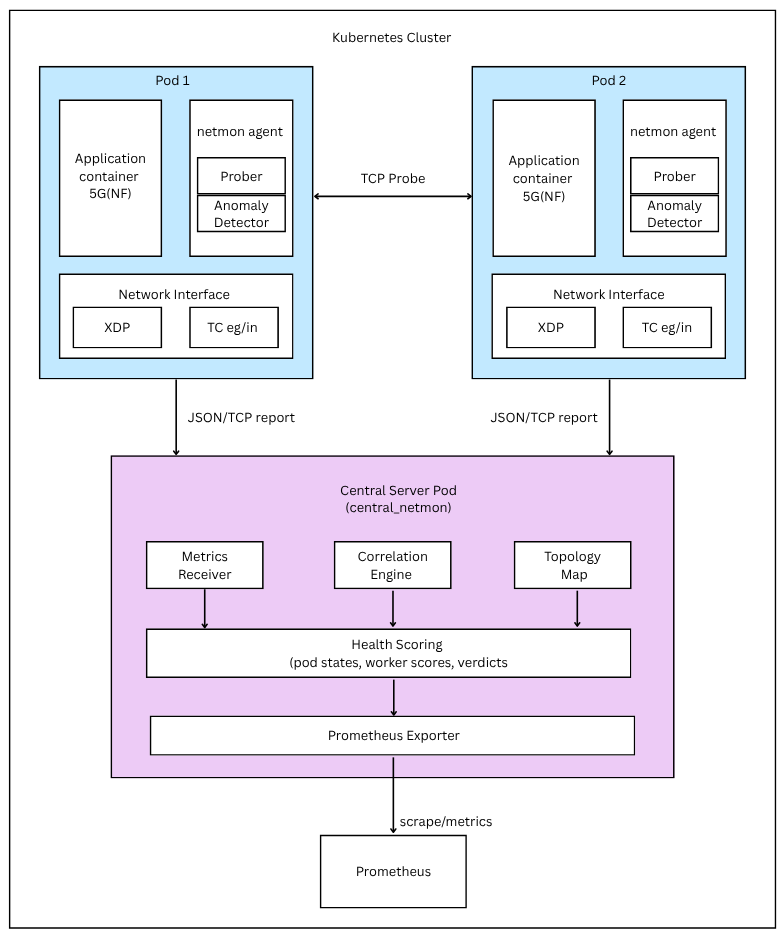}
    \caption{Architecture overview of centralized network monitoring for 5G pods in Kubernetes.}
    \label{fig:architecture overview}
\end{figure}

The \textbf{agent} actively probes all peers, passively monitors traffic via eBPF, performs local anomaly detection, and reports findings every 5 seconds via JSON over TCP. The \textbf{central server} aggregates reports, performs correlation analysis, maintains per-pair baselines using CUSUM change-point detection, maps cluster topology, and tracks pod and worker health states. Communication relies on Kubernetes service discovery via a headless service that resolves to current pod IPs~\cite{kubernetes_services}, enabling automatic adaptation as pods are added, removed, or rescheduled~\cite{kubernetes_lifecycle}.

\subsection{Distributed Monitoring Agent}
\label{sec:monitoring-agent}

To gather high-fidelity network telemetry directly from the data plane, the NetMon agent runs as a sidecar container that monitors its local pod environment using three primary mechanisms: active probing, passive eBPF monitoring, and local anomaly detection.

\subsubsection{Active Probing:}

The monitoring agents resolve the headless service name during each probe cycle to discover active peer pods within the cluster~\cite{kubernetes_services}. Every 5 seconds, each agent probes each peer via a TCP connection with a 3-second timeout. The probe uses a ping-pong protocol that exchanges worker node names, verifying connectivity and enabling topology discovery. Failures are tracked per peer: each failure generates a \texttt{PROBE\_FAILURE} anomaly, and $\geq$3 consecutive failures generate \texttt{CONSECUTIVE\_FAILURES}. A 60-second peer grace period ensures that terminated pods are detected via probe failures rather than silently disappearing from the monitoring set.

\subsubsection{eBPF Passive Monitoring:}

To capture network behavior directly from the Linux kernel without incurring userspace scheduling overhead, three custom eBPF programs are attached to the pod's primary network interface.

\textbf{eXpress Data Path(XDP) Program:}
This program runs at the earliest ingress point~\cite{hoiland_jorgensen_xdp_2018}, inspecting packets from known peers. It maintains per-peer counters (TCP/UDP packets, bytes, per-flag counts, retransmissions) in BPF maps. Potential retransmission detection uses an LRU hash of (source IP, source port, TCP sequence number) with a 50ms minimum age floor to avoid false positives from kernel packet batching. RST and retransmission events are forwarded via a ring buffer for userspace processing.

\textbf{TC Programs:}
A pair of egress and ingress programs measure the end-to-end probe latency. The egress program timestamps outgoing SYN packets destined for the probe port, while the ingress program calculates latency when the first data-bearing reply segment arrives from the peer, emitting the RTT (in microseconds) via a ring buffer. This captures the full round-trip from connection initiation to application response, including network traversal, TCP handshake completion, and remote processing time. Because timestamping occurs in the kernel's TC layer, local userspace scheduling between send and receive does not affect the measurement, though the remote peer's application processing time is included. The TC egress program additionally tracks per-flow traffic counters for cluster-wide application traffic analysis.

\subsubsection{Local Anomaly Detection:}

The agent processes the kernel metrics collected using two main local detection strategies.

\textbf{Z-score detection.}
This is the first strategy used to analyze latency metrics. For latency metrics, an Exponential Moving Average (EMA, $\alpha = 0.2$)~\cite{heckert2002handbook} serves as the adaptive baseline. The running standard deviation is computed as the square root of an EMA of squared deviations ($\alpha_{\text{var}} = 0.05$) with a minimum floor of 0.5ms. An anomaly is raised when the z-score exceeds 5.0 and the absolute deviation is $\geq$2ms. Detection requires $\geq$20 samples and is performed \emph{before} updating the EMA and variance, ensuring that the z-score reflects deviation from the pre-existing baseline.

\textbf{Relative and absolute thresholds.}
This is the second strategy designed to monitor traffic anomalies. For traffic statistics, RST and retransmission rates are flagged when exceeding $5\times$ their EMA baseline or absolute thresholds of 5\% (RST) and 10\% (retransmissions).



Using these strategies, the agent evaluates the incoming data points to detect and report the following concrete anomaly types:

\begin{itemize}
\item The agent triggers \texttt{KERNEL\_RTT\_SPIKE} and \texttt{JITTER\_SPIKE} anomalies when the z-score of the round-trip time or jitter exceeds 5.0 with an absolute deviation of at least 2ms.
\item It reports \texttt{PROBE\_FAILURE} and \texttt{CONSECUTIVE\_FAILURES} anomalies to track immediate or sustained active connection failures between peer pods.
\item It flags \texttt{HIGH\_RST} and \texttt{HIGH\_RETRANSMIT} anomalies when the absolute ratio of TCP resets exceeds 5\% or packet retransmissions exceed 10\%.
\item It raises \texttt{RST\_SPIKE} and \texttt{RETX\_SPIKE} anomalies when the short-term rate of resets or retransmissions exceeds five times the historical EMA baseline.
\item It registers a \texttt{HIGH\_PPS} anomaly when the incoming packet-per-second rate spikes to more than five times its expected historical baseline.
\item It generates a \texttt{TRAFFIC\_SILENCE} anomaly when an active peer pod fails to send any detectable network traffic for more than 30 seconds.
\end{itemize}

\subsection{Central Server Design}
The central server aggregates reports from all agents, performs cluster-wide correlation, maintains topology information, and tracks pod and worker health states.

\subsubsection{Pod State and Health Classification:}
The central server maintains per-pod state including lifetime metrics, current peer states, and health classification determined by counting probe failures and anomalies \emph{targeting} the pod within a 30-second window:

\begin{itemize}
    \item \texttt{DOWN}: no report for $>60$s
    \item \texttt{FAILING}: $\geq$9 probe failures
    \item \texttt{WARNING}: $\geq$3 failures or $\geq$30 anomalies
    \item \texttt{DEGRADED}: $\geq$1 failure or $\geq$20 anomalies
    \item \texttt{HEALTHY}: none of the above
\end{itemize}

The health state is based on what other pods report \emph{about} a pod, which enables detection of problems that are invisible to the pod itself. The precedence of the health states are sequential, so a pod that is both \texttt{WARNING} and \texttt{FAILING}, would show up as \texttt{FAILING} only.

\subsubsection{Correlation Engine:}
The correlation engine analyzes anomaly history within a 30-second sliding window, recomputed on every incoming report. Nine correlation patterns are identified in three categories.

\textbf{Pod-level correlations:}
\begin{itemize}
    \item \texttt{PEER\_DOWN}: $\geq$9 probe failures target the same peer
    \item \texttt{LATENCY\_SOURCE}: $\geq$15 RTT/jitter anomalies target the same pod
    \item \texttt{ANOMALY\_HOTSPOT}: $\geq$30 anomalies target the same pod
\end{itemize}

\textbf{Connectivity correlations:}
\begin{itemize}
    \item \texttt{WORKER\_ISOLATED}: pod fails to reach $\geq$3 peers on the same worker
    \item \texttt{INTER\_NODE\_FAILURE}: worker link $<50\%$ success while other links remain healthy
    \item \texttt{POD\_ISOLATED}: pod $<10\%$ success rate while other pods on same worker remain healthy
\end{itemize}

\textbf{Worker-level correlations:}
\begin{itemize}
    \item \texttt{WORKER\_HOTSPOT}: $\geq$2 external workers report anomalies and health score is $\geq10$ below average
    \item \texttt{WORKER\_HIGH\_LATENCY}: $>50\%$ of inter-worker links have RTT $>10$ms
    \item \texttt{WORKER\_SOURCE\_HOTSPOT}: worker reports $\geq40$ anomalies about $\geq3$ others and $\geq2\times$ the cluster average
\end{itemize}

\subsubsection{Online Change-Point Detection (CUSUM):}
\label{sec:cusum}
For each source-destination pod pair, the central server maintains running mean and variance using Welford's online algorithm (numerically stable, constant memory per metric). CUSUM (Cumulative Sum) detects distributional shifts by accumulating standardized deviations from the learned mean, with slack parameter $k=0.5$ controlling sensitivity and decision threshold $h=4.0$. Upon detection, the accumulators reset and the mean updates to the current value, adapting to the new operating point. A warmup of 30 samples and a minimum 5ms (for RTT only) absolute deviation are required.

The system generates three CUSUM-based anomalies (\texttt{CUSUM\_RTT\_SHIFT}, \texttt{CUSUM\_JITTER\_SHIFT}, and \texttt{CUSUM\_RETX\_SHIFT}).

\subsubsection{Topology Awareness and Worker Health:}
\label{sec:topology_awareness}

Worker node assignments obtained through probe handshakes enable topology-aware analysis. Each worker receives a health score combining probe success rate (weight 0.6) and anomaly rate (weight 0.4) within a 30-second window, expressed on a 0--100 scale.

Traffic baselines between worker pairs are also maintained to detect traffic drops ($<40\%$ of baseline) and traffic spikes ($>3\times$ baseline).

\subsection{Reporting and Visualization}

The server exposes a Prometheus endpoint with metrics including pod health states, per-peer RTT, jitter, throughput, worker health scores, CUSUM statistics, and active correlations. These metrics are scraped and visualized through Ericsson's CNOM (Cloud Native Operations Manager) dashboard platform.

\section{Implementation and Deployment}

The system is implemented in C using \texttt{libbpf} for eBPF program loading. eBPF programs are compiled with Clang targeting BPF and loaded via the skeleton mechanism. Deployment requires \texttt{CAP\_BPF} and \texttt{CAP\_NET\_ADMIN}~\cite{noauthor_capabilities7_nodate}. Key trade-offs: eBPF over packet capture (lower overhead, requires Linux 5.8+), push-based reporting (lower latency, single point of failure), $O(N^2)$ full-mesh probing (complete coverage, limits scalability), sensitivity over precision (mitigated by multi-source correlation), statistical methods over ML (no training data needed, interpretable).

\subsection{Evaluation Methodology}

All experiments are conducted on a live Kubernetes cluster since live tests capture realistic kernel and network behavior.

\textbf{Environment:}
\label{sec:test_cluster}
The test environment consists of a virtual 5G AMF deployed across four worker nodes with ten pods spanning four microservice types (controller, SCTP transport, forwarding, mobility management), with 2--3 pods per worker.

\begin{figure}[htbp]
    \centering
    \includegraphics[width=0.8\linewidth]{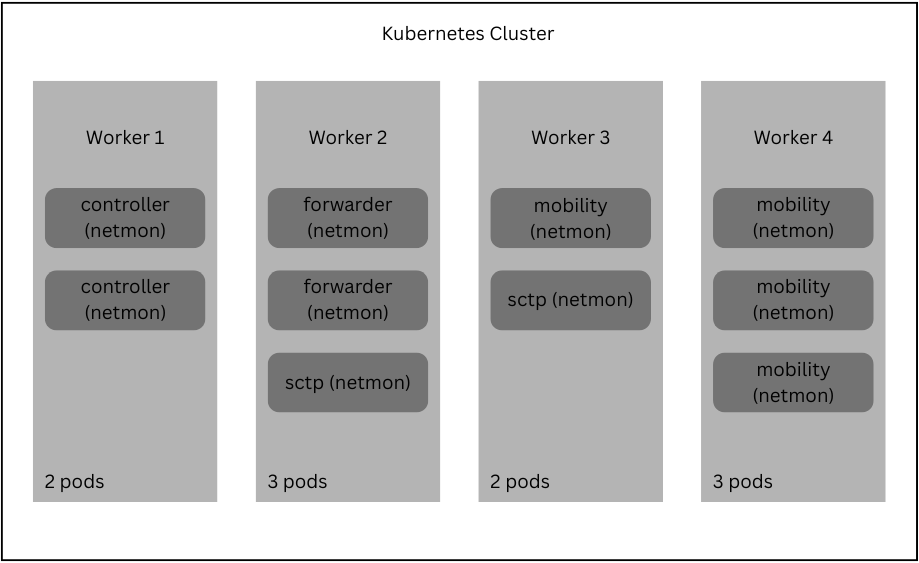}
    \caption{Cluster topology with 10 pods across 4 worker nodes.}
    \label{fig:test_topology}
\end{figure}

\textbf{Fault Injection:}
Controlled faults were injected using \texttt{tc} and \texttt{iptables} on a single target worker, with the central server on a separate worker. Each scenario was repeated $\geq$5 times at 0\% simulated UE load, and 2 times at 20\%, and 50\% simulated UE load. Injections lasted 2 minutes with 5-minute recovery periods. Table \ref{tab:fault-scenarios} summarizes the configurations of each scenario of fault injection.

\textbf{Metrics:} 
To evaluate the performance, efficiency, and diagnostic accuracy of NetMon under various fault injection scenarios, we track three primary categories of metrics:
\begin{itemize}
    \item \textbf{Detection latency}: Time from fault injection to the first relevant anomaly observed at the central server involving the affected worker.
    \item \textbf{Correlation latency}: Time to the first worker-level correlation that correctly localizes the fault.
    \item \textbf{Resource overhead}: CPU usage (\texttt{kubectl top} and \texttt{bpftool} statistics), memory usage (RSS), and network traffic ($N(N-1)$ probes and $N$ reports per cycle).
\end{itemize}

\vspace{-1cm}

\begin{table}[htbp]
\centering
\caption{Fault injection scenarios and configurations.}
\label{tab:fault-scenarios}
\renewcommand{\arraystretch}{1.25}
\footnotesize
\begin{tabular}{|p{2cm}|p{3.4cm}|p{3.0cm}|p{3.8cm}|}
\hline
\textbf{Scenario} & \textbf{Method} & \textbf{Configurations} & \textbf{Target Detection} \\
\hline
\scriptsize{Latency} & \scriptsize{tc netem delay} & \scriptsize{10--100 ms, 0/20/50\%} & \tiny{KERNEL\_RTT\_SPIKE}, \tiny{WORKER\_HIGH\_LATENCY}\\
\hline
\scriptsize{Jitter} & \scriptsize{tc netem delay 5 ms $\pm$20 ms} & \scriptsize{0/20/50\%} & \tiny{JITTER\_SPIKE}, \tiny{WORKER\_HIGH\_LATENCY} \\
\hline
\scriptsize{Packet} loss & \scriptsize{tc netem loss} & \scriptsize{5--50\%, 0/20/50\%} & \tiny{RETX\_SPIKE}, \tiny{WORKER\_SOURCE\_HOTSPOT}\\
\hline
\scriptsize{Bandwidth} & \scriptsize{tc tbf rate} & \scriptsize{1 and 5 Mbit/s, 0\%} & \tiny{WORKER\_SOURCE\_HOTSPOT} \\
\hline
\scriptsize{Partition} & \scriptsize{iptables} & \scriptsize{Full isolation, 0\%} & \tiny{PEER\_DOWN}, \tiny{WORKER\_ISOLATED} \\
\hline
\scriptsize{Link failure} & \scriptsize{iptables} & \scriptsize{Worker pair, 0\%} & \tiny{PEER\_DOWN}, \tiny{WORKER\_ISOLATED} \\
\hline
\scriptsize{Pod termination} & \scriptsize{kubectl delete} & \scriptsize{Single pod, 0\%} & \tiny{PROBE\_FAILURE}, \tiny{PEER\_DOWN} \\
\hline
\scriptsize{Gradual} & \scriptsize{tc netem} stepped & \scriptsize{1 and 5 ms steps, 0/20/50\%} & \tiny{CUSUM\_RTT\_SHIFT}, \tiny{WORKER\_HIGH\_LATENCY} \\
\hline
\end{tabular}
\end{table}

\vspace{-1cm}

\section{Evaluation}

\subsection{CNOM Dashboard Visualization}

\begin{figure}[h]
    \centering
    \includegraphics[width=1\linewidth]{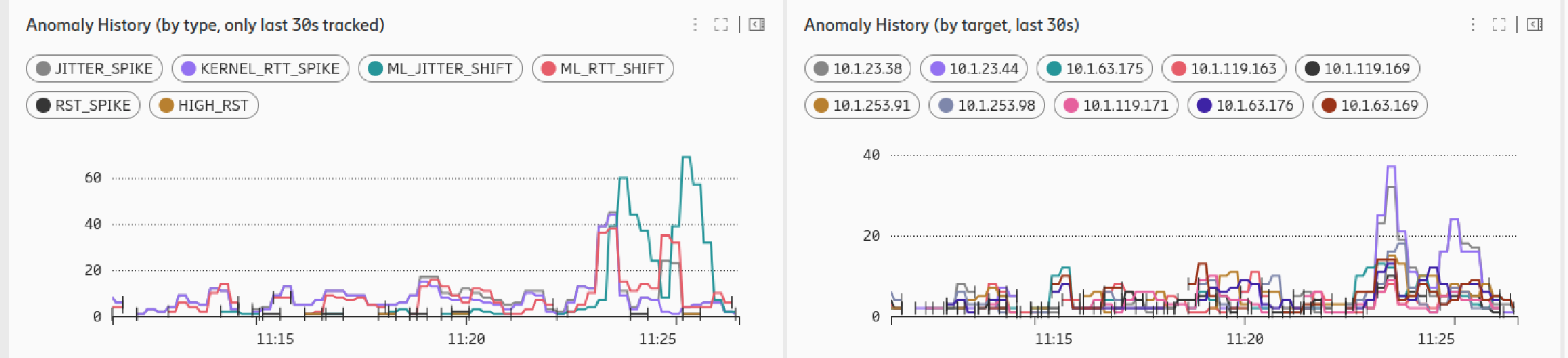}
    \includegraphics[width=1\linewidth]{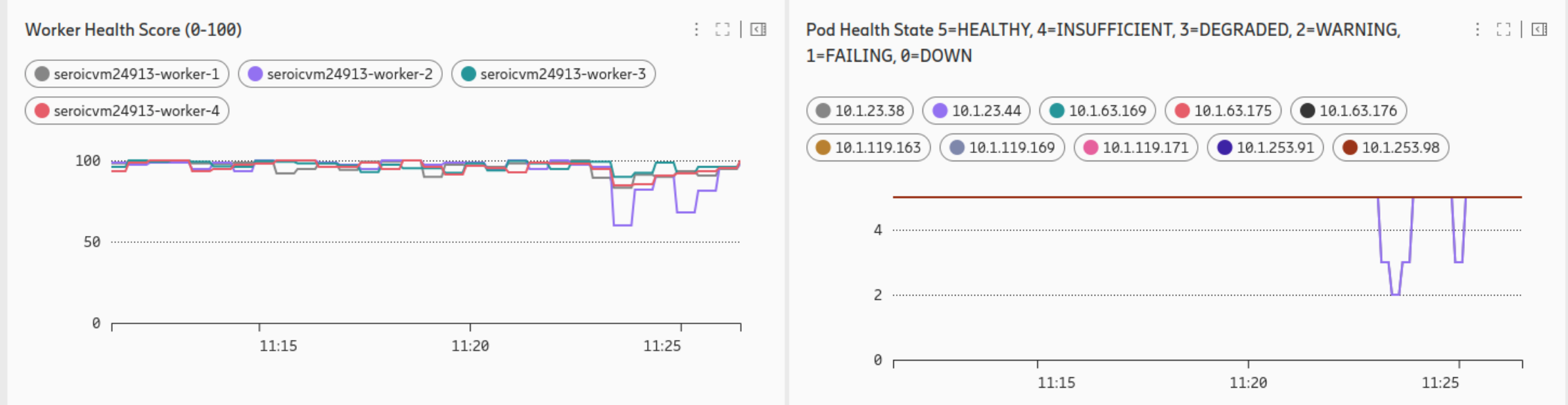}
    \begin{subfigure}[h]{0.50\linewidth}
        \includegraphics[width=\linewidth]{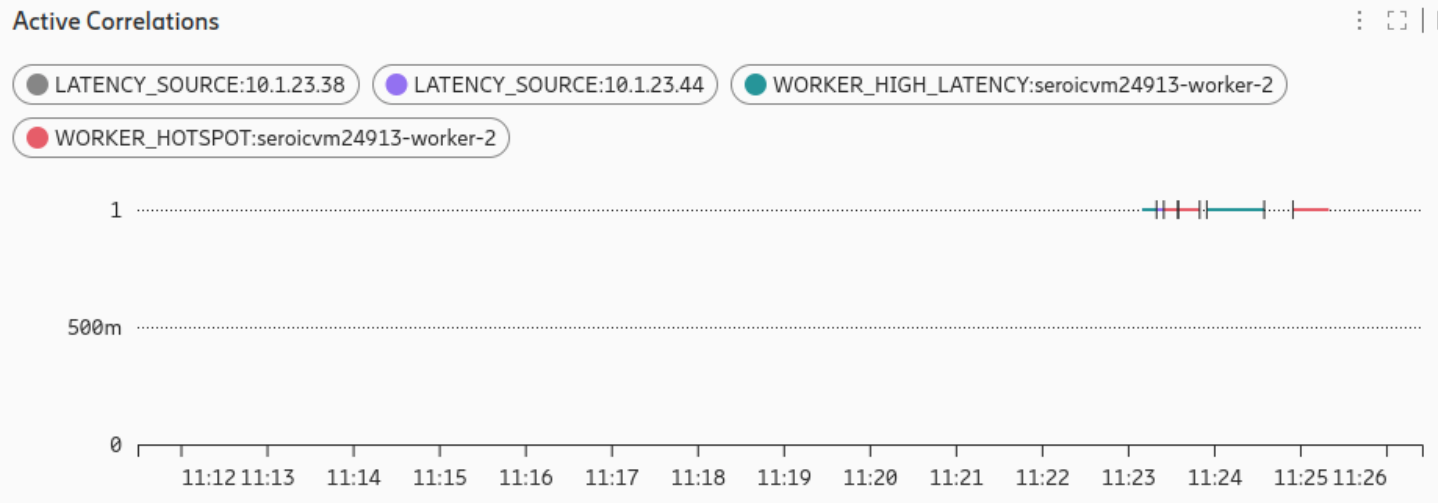}    
    \end{subfigure}
    \begin{subfigure}[h]{0.49\linewidth}
        \includegraphics[width=\linewidth]{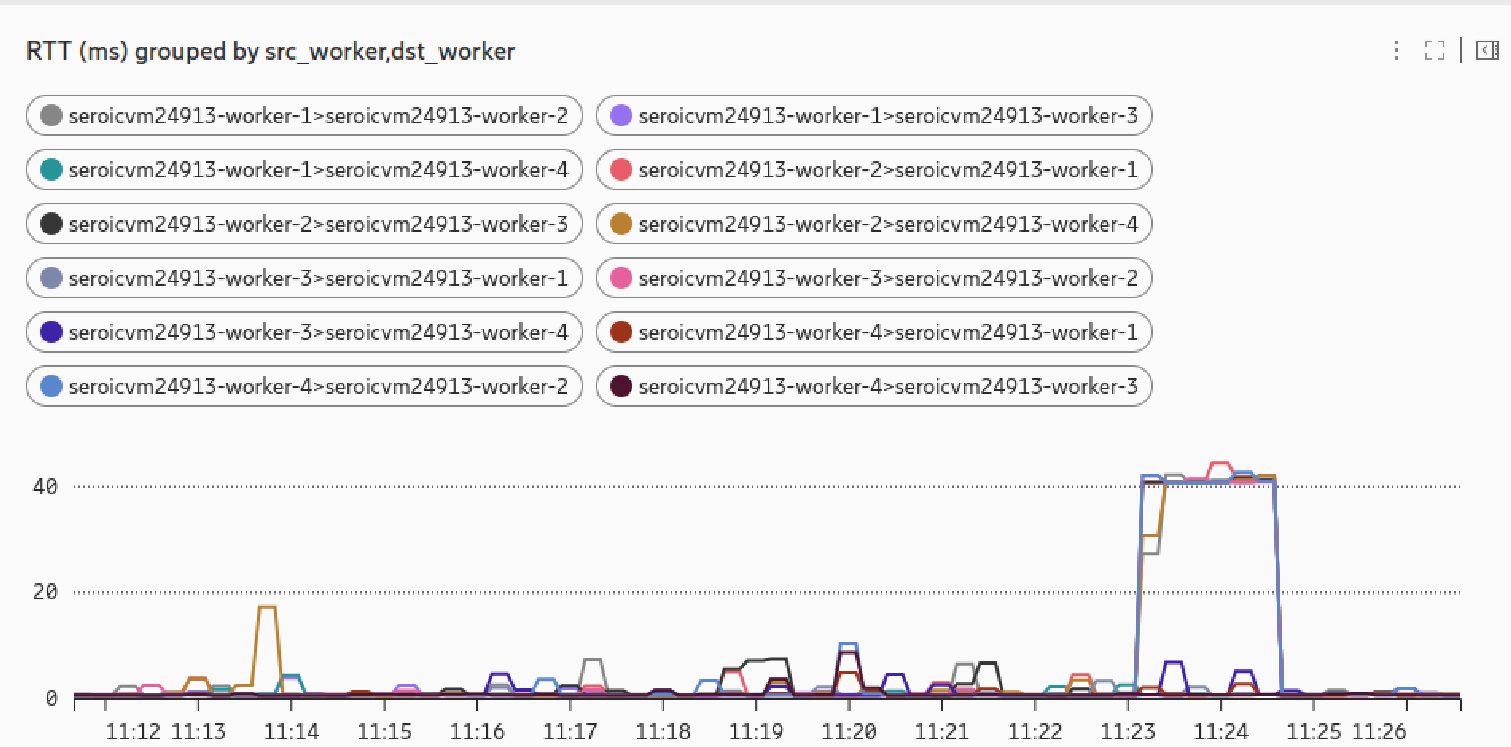}    
    \end{subfigure}
    \caption{CNOM dashboard widgets during 20ms latency injection showing anomaly detection, correlation, and health state response.}
    \label{fig:cnom-dash}
\end{figure}

Figure~\ref{fig:cnom-dash} shows the dashboard during a 20ms latency injection on worker-2: latency spikes (bottom right) generate anomalies (top two widgets), which trigger \newline\texttt{WORKER\_HIGH\_LATENCY} and \texttt{WORKER\_HOTSPOT} correlations (bottom-left), causing worker health scores and pod states to drop (middle two widgets), all returning to baseline after fault removal. The two-level presentation, raw anomalies and correlations, allows operators to identify faults without manually inspecting per-pod metrics.

\subsection{Resource Overhead}
\label{sec:cpu-mem-usage}
Resource consumption is approximately constant across all tested conditions. Each agent consumes 3.4 millicores (3m userspace + 0.4m eBPF kernel-side) and 4.5~MiB memory. The central correlator uses <1 millicore and 2.6~MiB. For 10 pods, total overhead is \textless 35 millicores and 47.6~MiB, representing 0.5--8.7\% of application CPU and 0.5--0.9\% of application memory depending on simulated load (lower load = higher percentage). The resulting network overhead comprises 90 active probes of less than 200~bytes each and 10 metrics reports of approximately 4--5~KB each, which are transmitted every 5~seconds.

\subsection{Baseline Behavior}

Table~\ref{tab:baseline} characterizes the system during 18 hours of normal operation (0\% load, no faults).

\begin{table}[htbp]
\centering
\caption{Anomaly and correlation rates during normal operation (18 hours, no faults).}
\label{tab:baseline}
\small
\begin{tabular}{|l|r|r|}
\hline
\textbf{Metric} & \textbf{Total} & \textbf{Rate (/hr)} \\
\hline
CUSUM\_RTT\_SHIFT & 16156 & 893 \\
JITTER\_SPIKE & 15812 & 874 \\
KERNEL\_RTT\_SPIKE & 14383 & 795 \\
CUSUM\_JITTER\_SHIFT & 3484 & 192 \\
RST\_SPIKE & 1628 & 90 \\
HIGH\_RST & 1051 & 58 \\
\hline
\textbf{Total anomalies} & \textbf{52514} & \textbf{2902} \\
\hline
WORKER\_HOTSPOT & 4084 & 225 \\
WORKER\_SOURCE\_HOTSPOT & 2770 & 153 \\
LATENCY\_SOURCE & 563 & 31 \\
WORKER\_HIGH\_LATENCY & 197 & 10 \\
ANOMALY\_HOTSPOT & 38 & 2 \\
PEER\_DOWN / WORKER\_ISOLATED & 0 & 0 \\
\hline
\textbf{Total correlations} & \textbf{7652} & \textbf{423} \\
\hline
\end{tabular}
\end{table}

The observed \(\sim\)2900 anomalies per hour reflect real deviations of the RTT in the virtualized environment. The hard-failure correlations \texttt{PEER\_DOWN} and \texttt{WORKER\_ISOLATED} produced zero activations during baseline measurements, providing unambiguous signals when they fire. Reducing anomaly volume would require raising detection thresholds and increasing detection latency, a deliberate trade-off favoring sensitivity. The observed 423 correlations per hour are dominated by \texttt{WORKER\_HOTSPOT} and \texttt{WORKER\_SOURCE\_HOTSPOT} which are generated purely from the amount of anomalies generated. So while these two correlations provide some information about cluster stability, they should be paired with other signals in order to more accurately diagnose faults.

\subsection{Experiment Results}
Six fault categories were evaluated at 0\%, 20\%, and 50\% load. Only anomalies concerning the injected worker count toward detection latency.

\subsubsection{Latency Injection (Table \ref{tab:latency-detection}):} Detection within 0--10s across all magnitudes. Correlation improves under load (6--10s at 50\% vs 13--20s at 0\%) because elevated baseline RTT pushes measurements further above the 10ms threshold.
\vspace{-0.5cm}

\begin{table}[h]
\centering
\caption{Detection and correlation latency for latency injection (seconds). Correlation = \texttt{WORKER\_HIGH\_LATENCY}. R1-5=0\%, R6-7=20\%, R8-9=50\%}
\label{tab:latency-detection}
\small
\begin{tabular}{|l|ccccc|cc|cc|}
\hline
& \multicolumn{5}{c|}{0\% load} & \multicolumn{2}{c|}{20\%} & \multicolumn{2}{c|}{50\%} \\
\textbf{Delay} & R1 & R2 & R3 & R4 & R5 & R6 & R7 & R8 & R9 \\
\hline
\multicolumn{10}{|l|}{\textit{Detection latency}} \\
\hline
10ms  & 10 & 2 & 7 & 6 & 5 & 2 & 4 & 2 & 2 \\
25ms  & 3  & 9 & 0 & 7 & 3 & 4 & 3 & 2 & 7 \\
50ms  & 0  & 5 & 3 & 3 & 7 & 3 & 3 & 3 & 3 \\
100ms & 4  & 3 & 3 & 3 & 3 & 5 & 2 & 3 & 3 \\
\hline
\multicolumn{10}{|l|}{\textit{Correlation latency}} \\
\hline
10ms  & 20 & 19 & 17 & 20 & 17 & 14 & 18 & 10 & 10 \\
25ms  & 13 & 15 & 19 & 17 & 16 & 14 & 14 & 8  & 10 \\
50ms  & 15 & 18 & 15 & 18 & 16 & 7  & 15 & 8  & 8  \\
100ms & 14 & 13 & 14 & 13 & 13 & 11 & 4  & 6  & 6  \\
\hline
\end{tabular}
\end{table}

\subsubsection{Jitter Injection (Table \ref{tab:jitter-detection}):}
Detection within 0--5s; correlation within 6--19s correctly identifying the affected worker in all runs.
\vspace{-0.5cm}

\begin{table}[H]
\centering
\caption{Detection and correlation latency for jitter injection (5ms $\pm$ 20ms). All values in seconds.}
\label{tab:jitter-detection}
\small
\begin{tabular}{|l|ccccc|cc|cc|}
\hline
& \multicolumn{5}{c|}{0\% load} & \multicolumn{2}{c|}{20\%} & \multicolumn{2}{c|}{50\%} \\
& R1 & R2 & R3 & R4 & R5 & R6 & R7 & R8 & R9 \\
\hline
Detection (s)   & 5  & 1 & 2 & 2 & 2 & 0 & 5 & 0 & 1 \\
Correlation (s) & 15 & 18 & 19 & 16 & 17 & 8 & 6 & 7 & 8 \\
\hline
\end{tabular}
\end{table}
\vspace{-0.5cm}

\subsubsection{Packet Loss (Table \ref{tab:loss-detection}):}

\vspace{-0.5cm}
\begin{table}[H]
\centering
\caption{Detection and correlation latency for packet loss (seconds). Correlation = \texttt{WORKER\_SOURCE\_HOTSPOT}; -- = did not fire (1/9 runs at 50\% loss).}
\label{tab:loss-detection}
\small
\begin{tabular}{|l|ccccc|cc|cc|}
\hline
& \multicolumn{5}{c|}{0\% load} & \multicolumn{2}{c|}{20\%} & \multicolumn{2}{c|}{50\%} \\
\textbf{Loss} & R1 & R2 & R3 & R4 & R5 & R6 & R7 & R8 & R9 \\
\hline
\multicolumn{10}{|l|}{\textit{Detection latency}} \\
\hline
5\%  & 7  & 7  & 7  & 10 & 5  & 11 & 10 & 10 & 6  \\
10\% & 4  & 16 & 5  & 10 & 12 & 3  & 6  & 4  & 5  \\
20\% & 7  & 9  & 8  & 13 & 12 & 6  & 17 & 7  & 5  \\
50\% & 22 & 58 & 19 & 17 & 26 & 19 & 14 & 6  & 17 \\
\hline
\multicolumn{10}{|l|}{\textit{Correlation latency}} \\
\hline
5\%  & 35 & 37 & 35 & 37 & 71 & 22 & 37 & 39 & 71 \\
10\% & 13 & 46 & 42 & 22 & 34 & 21 & 32 & 11 & 10 \\
20\% & 24 & 10 & 22 & 25 & 25 & 14 & 20 & 11 & 16 \\
50\% & 133 & 98 & 19 & 47 &-- & 24 & 129 & 24 & 43 \\
\hline
\end{tabular}
\end{table}
\vspace{-0.5cm}

Detection in 3--17s at 5--20\% loss, increasing to 6--58s at 50\% (fault drops metric reports). Correlation latency is higher than for latency faults due to the probabilistic nature of retransmission-based detection.

\subsubsection{Bandwidth Throttle, Link Failure, Pod Termination, and Gradual Degradation:}

Bandwidth throttling at 5~Mbit/s was detected in 0--9s with correlation in 7--27s. At 1~Mbit/s, metric report starvation reduced correlation reliability; experiments under load were invalid as the constraint was insufficient for both application and monitoring traffic.

Link failure (worker-2 $\leftrightarrow$ worker-3) detected in 5--9s, \texttt{WORKER\_ISOLATED} in 14--19s, \texttt{PEER\_DOWN} in 22--25s. Full network partition (worker-2 isolated): detected and correlated simultaneously in 9--11s, succeeding even when the partitioned side cannot report.

Pod termination detected in 3--9s (9/10 runs, one 29s outlier from extended graceful shutdown), \texttt{PEER\_DOWN} in 10--15s. A peer grace period ensures terminated pods remain monitored until confirmed unreachable.

Gradual Degradation With 5ms steps, detection and correlation within 2--9s. With 1ms steps applied every 30s, \texttt{WORKER\_HIGH\_LATENCY} fires consistently at the 5ms accumulated level (${\sim}$133s after injection start), representing the minimum detectable gradual degradation via correlation at thi step interval.

With 1ms steps, \texttt{WORKER\_HIGH\_LATENCY} fires consistently at the 5ms accumulated level (~133s), representing the minimum detectable gradual degradation via correlation.

\vspace{-0.8cm}

\begin{table}[H]
\centering
\caption{Detection and correlation latency for gradual latency increase (seconds).}
\label{tab:gradual-detection}
\small
\begin{tabular}{|l|ccccc|cc|cc|}
\hline
& \multicolumn{5}{c|}{0\% load} & \multicolumn{2}{c|}{20\%} & \multicolumn{2}{c|}{50\%} \\
\textbf{Step size} & R1 & R2 & R3 & R4 & R5 & R6 & R7 & R8 & R9 \\
\hline
\multicolumn{10}{|l|}{\textit{5ms steps (0--50ms)}} \\
\hline
Detection   & 2 & 4 & 2 & 2 & 7 & 5 & 4 & 5 & 5 \\
Correlation & 8 & 8 & 8 & 6 & 9 & 7 & 8 & 7 & 8 \\
\hline
\multicolumn{10}{|l|}{\textit{1ms steps (0--20ms)}} \\
\hline
Detection   & 4   & 1   & 2   & 2   & 0   & 7   & 8   & 6   & 1   \\
Correlation & 133 & 135 & 133 & 134 & 134 & 133 & 135 & 133 & 133 \\
\hline
\end{tabular}
\end{table}

\vspace{-1cm}

\subsection{Summary}

\begin{figure}[htbp]
    \centering
    \includegraphics[width=0.8\linewidth]{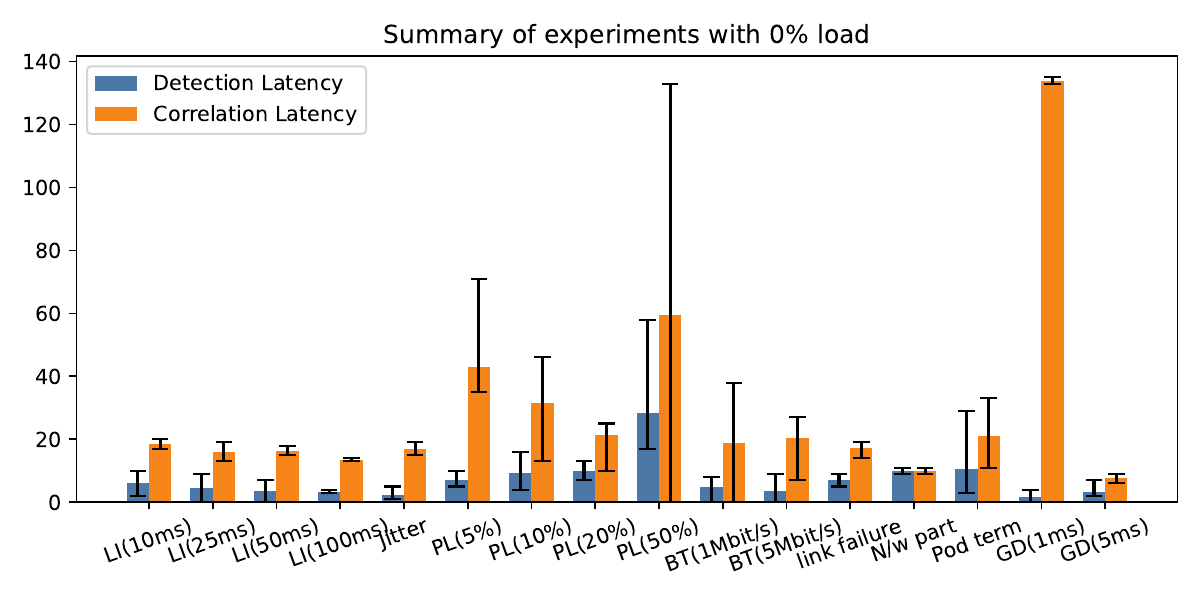}
    \caption{Detection and correlation latencies at 0\% load. LI: Latency Injection, PL, Packet Loss, BT: Bandwidth Throttle, N/w part: Network Partition, GD: Gradual Degradation.}
    \label{fig:summary-0pct}
\end{figure}

Figure~\ref{fig:summary-0pct} presents detection and correlation latencies across all fault types at 0\% load. Results at 20\% and 50\% load show equal or improved performance due to elevated baseline RTT pushing measurements further above correlation thresholds.

\vspace{-0.8cm}

\vspace{-0.5cm}

The system detects faults as subtle as 10ms latency or 5\% packet loss, localizes them to the affected worker within seconds, maintains this capability under 50\% load, and introduces negligible resource overhead. The main limitations are baseline anomaly volume in virtualized environments, self-limiting visibility under severe bandwidth constraints, and $O(N^2)$ probe scaling.

\section{Discussion and Limitations}

\subsection{Discussion}

\textbf{Detection and localization performance:}
The system detects latency and jitter faults within 2--10 seconds and localizes them to the correct worker within 6--20 seconds. In the tested configurations, this is comparable to or faster than typical Kubernetes liveness probe intervals of 10--30 seconds~\cite{kubernetes_liveness}.This addresses the ``gray failure'' scenario~\cite{huang2017gray} where components remain alive but perform poorly, a condition standard liveness probes cannot detect.

\textbf{Comparison with existing approaches:}
Prometheus with Blackbox Exporter~\cite{prometheus_blackbox} typically requires at least one scrape cycle plus alerting rule evaluation before firing an alert, with end-to-end latency depending on the configured scrape interval (commonly 15--60s~\cite{prometheus_config}) and evaluation frequency. Furthermore, it probes from an external vanatage point, lacking the internal data-plane perspective. Purely passive eBPF tools~\cite{soldani_ebpf_2023} cannot verify connectivity on idle paths, critical in 5G where traffic is bursty and silent partitions between low-traffic pods would go undetected. NetMon's active probing verifies every path every 5 seconds regardless of traffic patterns.

\textbf{Resource overhead:}
At 3.4 millicores CPU and 4.5~MiB memory per pod (<1\% of application resources at 50\% load), the system demonstrated low measured overhead in this setup during periods of stress when monitoring is most critical. Overhead remains constant regardless of load or fault state.

\textbf{Correlation accuracy:}
\texttt{PEER\_DOWN} and \texttt{WORKER\_ISOLATED} produced no false positives during the 18-hour baseline. Softer correlations such as \texttt{WORKER\_HOTSPOT} and \texttt{WORKER\_SOURCE\_HOTSPOT} provide useful indicators but require operator judgment. The full partition experiment correctly identified the fault within 10s using only observations from the non-partitioned side.

\textbf{Threshold sensitivity and portability:}
Detection thresholds (z-score: 5.0, CUSUM slack: 0.5, minimum deviation: 2ms, correlation triggers) were tuned iteratively on preliminary baseline runs rather than dedicated validation splits. While this iterative tuning was tailored for this environment, it carries a risk of overfitting to the specific test cluster's noise characteristics. Different infrastructure would likely require recalibration, a inherent property of static threshold-based detection that favors interpretability and operator control over automatic adaptation. The CUSUM component partially mitigates this limitation through online baseline learning.

\subsection{Limitations}
\label{sec:conc-limitations}

\begin{itemize}
    \item \textbf{Baseline noise:} The virtualized environment's 2--5ms RTT variance generates ~2900 anomalies/hour during normal operation, producing supplementary correlations requiring operator judgment. Bare-metal deployments would probably exhibit a lower noise floor.

    \item \textbf{$O(N^2)$ probe scaling:} Full-mesh probing is manageable at the evaluated cluster scale (10 pods), but $O(N^2)$ network and CPU overhead bounds its scalability in larger deployments, requiring probe sampling or hierarchical aggregation.

    \item \textbf{Single point of failure:} The central correlator is a single instance; its loss disables cluster-wide correlation until rescheduled, though agents continue local detection.

    \item \textbf{Bandwidth throttle visibility:} Severe throttling starves metric reports, reducing visibility when monitoring is most needed, a limitation inherent to any in-band monitoring approach.

    \item \textbf{Elevated capabilities:} eBPF requires \texttt{CAP\_BPF} and \texttt{CAP\_NET\_ADMIN}, which may conflict with hardened production security policies.

    \item \textbf{Visualization constraints:} CNOM limits time-series to 20 lines per widget and offers fewer capabilities than Grafana. The detection engine is independent of the visualization layer and could be paired with a more capable frontend.
\end{itemize}

\section{Conclusion and Future Work}

\subsection{Conclusion}

This paper presented a hybrid network monitoring system for Kubernetes-based 5G deployments combining eBPF passive observation, active TCP probing, and centralized correlation. Evaluation on a live AMF deployment demonstrated detection of faults as subtle as 10ms latency or 5\% packet loss within 2--10 seconds, correct localization within 6--20 seconds, and zero false positives for high-confidence correlations across 18 hours. Full network partitions are detected and correlated within 9--11 seconds. Total overhead of 3.4 millicores CPU and 4.5~MiB per pod suggests that the approach is feasible at scale in production.

The hybrid approach addresses a gap where standard health checks cannot detect partial degradation, scrape-based systems introduce tens-of-seconds delays, and passive tools cannot verify idle paths. By combining these complementary techniques, the system provides early detection and fault localization for maintaining service quality in cloud-native 5G infrastructure.

\subsection{Future Work}

\begin{itemize}
    \item \textbf{Expanded detection patterns:} Asymmetric latency detection, traffic flow correlation with probe failures, and temporal pattern recognition for recurring degradation.

    \item \textbf{Adaptive threshold learning:} Inferring thresholds from observed baseline variance per environment, reducing manual tuning when deploying to new clusters.

    \item \textbf{ML/AI integration:} Neural networks for complex baseline modeling, classification models to distinguish fault-induced correlations from noise, and LLM-based root cause summarization.

    \item \textbf{Probe scaling:} Hierarchical per-worker aggregation changing scaling from $O(N^2)$ to $O(W^2)$ where $W$ is the worker count, requiring per-worker leader election.

    \item \textbf{Central server fault tolerance:} Leader election among agents to remove dependency on a dedicated controller pod.
\end{itemize}

%
%
%
\bibliographystyle{splncs04}
\bibliography{references}
%


\end{document}